# Single-Cell Trajectory Reconstruction Reveals Migration Potential of Cell Populations

Yanping Liu[1,*], Dui Qin[1], Xinwei Li[1], Guoqiang Li[2], Zhichao Liu[1], Kena Song[3], Wei Wang[1], Zhangyong Li[1,*]

[1] Department of Biomedical Engineering, Chongqing University of Posts and Telecommunications, Chongqing 400065, China
[2] Chongqing Key Laboratory of Environmental Materials and Remediation Technologies, College of Chemistry and Environmental Engineering, Chongqing University of Arts and Sciences, Chongqing 402160, China
[3] College of Medical Technology and Engineering, Henan University of Science and Technology, Luoyang 471023, China
*Corresponding authors: liuyp@cqupt.edu.cn, lizy@cqupt.edu.cn

**Abstract**
Cell migration, which is strictly regulated by intracellular and extracellular cues, is crucial for normal physiological processes and the progression of certain diseases. However, there is a lack of an efficient approach to analyze super-statistical and time-varying characteristics of cell migration based on single trajectories. Here, we propose an approach to reconstruct single-cell trajectories, which incorporates wavelet transform, power spectrum of an OU-process, and fits of the power spectrum to analyze statistical and time-varying properties of customized target-finding and migration metrics. Our results reveal diverse relationships between motility parameters and dynamic metrics, especially the existence of an optimal parameter range. Moreover, the analysis reveals that the loss of Arpin protein enhances the migration potential of *D. discoideum*, and a previously reported result that the rescued amoeba is distinguishable from the wild-type amoeba. Significantly, time-varying dynamic metrics emerge periodic phenomena under the influence of irregularly changing parameters, which correlates with migration potential. Our analysis suggests that the approach provides a powerful tool for estimating time-dependent migration potential and statistical features of single-cell trajectories, enabling a better understanding of the relationship between intracellular proteins and cellular behaviors. This also provides more insights on the migration dynamics of single cells and cell populations.

## Introduction

Cell migration is an essential function that determines life and death events [1] and is indispensable for the normal development of tissues and organs, such as wound healing [2], morphogenesis [3], and immune responses [4]. In cancer progression, cells become dysregulated and can migrate away from the primary tumor site through the lymphatic or blood vessels to distant sites, a process known as metastasis [5].

In general, cell migration, characterized by two motility parameters [6], *i.e.*, persistence time $P$ and migration speed $S$, is regulated by intracellular signaling pathways [7] and extracellular microenvironments [8]. To elucidate the dynamic mechanisms underlying cell migration, a number of experiments are carried out, and some interesting and crucial phenomena emerge therewith [9]. For example, Han *et al*. constructed a Collagen-I Matrigel composite extracellular matrix (ECM) and found that locally aligned fibers can guide metastatic MDA-MB-231 breast cancer cells to invade rigid Matrigel (~10 mg/ml), implying that fiber alignment can be viewed as a pathway to enhance cell-ECM interactions [10]. Moreover, a quasi-3D in vitro model demonstrates that reorganized fiber bundles can carry tensile forces and guide strongly correlated migration [11]. Additionally, cell-ECM mechanical coupling also plays a crucial role in inducing spreading and aggregation in multi-cellular systems [12].

To better understand migration behaviors from individual to collective cells, some dynamic models have been constructed to intrinsically clarify potential principles [13-15]. For instance, more than thirty years ago, a model termed "persistent random walk" (PRW) was derived from a differential equation that describes the migration of a self-propelled cell, and it is commonly used to analyze the random migration of cells on 2D substrates [16]. Interestingly, the PRW model possesses several assumptions, including Gaussian distribution of migration velocity, single-exponential decay for auto-covariance function, isotropic migration ability, *etc*. Subsequently, Wu *et al*. found that 3D migration does not follow a random walk and thus proposed an anisotropic PRW (APRW) model to explain the effect of local anisotropy of microenvironments on cell migration [17]. Further, inspired by the models above, we take into account time-dependent properties of migration behaviors or/and microenvironments, and develop a time-varying PRW (TPRW) model to analyze abnormal acceleration profiles [18], non-linear velocity auto-correlation function (or follows a double-exponential decay) [19] and derive motility parameters versus time in previous works [20].

Additionally, researchers also constructed plentiful efficient approaches to uncover novel characteristics and modes of cell migration [21,22]. For example, mean-squared displacement (MSD) is developed to measure the migration ability of cells, bacteria, active particles and *etc.*, which enables one to directly evaluate the similarity of migration behaviors to ballistic ($\gamma$=2) or pure diffusive motility ($\gamma$=1) according to the slopes $\gamma$ of the MSD profiles [23,24]. Likewise, the velocity auto-covariance function (VAC) is a widely used measure to analyze the correlation of one migration velocity with another, meaning that one could determine the directional persistence of migration based on the VAC profiles [25,26]. Here, the persistence is also typically characterized by another metric, *i.e.*, the ratio of displacement to distance for individual trajectories [27]. To explore more information, the Fourier transform of VAC is performed to obtain the Fourier power spectrum (FPS) of migration velocity, which vividly shows how the spectral values are distributed with frequency [19,28]. Since the low (high) frequency corresponds to correlation on a long time scale (random noise on a short time scale), one can quickly determine the strength of persistence according to the spectral values [29].

More importantly, Gautreau *et al*. introduce a custom-made open-source computer program, *DiPer*, to quantify directional persistence in cell migration [30]. Considering the heterogeneous properties of extracellular environments, Fabry *et al*. combine an autoregressive process of first order (AR-1) with sequential Bayesian inference to extract persistence parameter $q$ and activity parameter $a$, and discriminate migration strategies in different environments [20]. Moreover, we also proposed an entropy-based approach that involves cellular turning dynamics and Shannon entropy to reflect the randomness or order of cell migration, with a value of 1 representing the most random diffusive migration and 0 representing the most ordered ballistic dynamics [31]. With a similar physical interpretation, a morphological entropy method is developed, enabling us to reveal migration mechanisms encoded in cellular morphology on multiple length scales, *i.e.*, cellular nucleus, single cell, and cell spheroid [32].

In this paper, we propose a novel approach called "single-cell trajectory reconstruction" (SCTR), which mainly involves in wavelet power spectrum of migration velocity, fits with power spectrum of OU-process, trajectory simulations via PRW model and calculations of dynamic metrics, and enables us to analyze migration potential of individual and population cells, especially including super-statistical and time-dependent characteristics. We first investigated the roles of motility parameters (persistence time and migration speed) in regulating dynamic metrics (target-finding and migration) and found the two time-varying metrics behave differently as the parameters change. Subsequently, the changing trends of the metrics are validated qualitatively by three groups of *D. discoideum* migration data (*i.e.*, wild-type amoeba, WT; knock-out amoeba, KO; rescued amoeba, RESCUE). Furthermore, we also reveal the differences between the WT (or RESCUE) and KO groups based on statistical features and, in particular, capture a previously unreported and significant discrepancy between the WT and RESCUE groups. In addition, we study the evolution of the metrics in detail and observe that periodic behaviors emerge in the time-varying metric profiles, which may indicate a stronger migration potential. Therefore, with this approach, we can analyze the super-statistical and time-dependent features and further elucidate migration potentials of individual cells as well as cell populations, such as cell migration regulated by intracellular proteins.

**Results**

**Real-time trajectory reconstruction of single-cell migration.**
In this paper, we develop an approach to decompose the velocity series of single-cell migration and further investigate the corresponding migration dynamics at each time point. We first extract the experimental or simulated position coordinates $r_i$ of a cell migrating in 3D or on 2D environments and calculate "momentary" velocities by applying the formula $v_i = (r_{i+1} - r_i)/\Delta t$ (Fig. 1A). Here the $\Delta t$ is the sampling time, which equals to the time lag between any two successive frames experimentally or the time interval used in computer simulations.

On the basis of migration velocities $v_i$, wavelet transform is introduced to compute the power spectrum of each velocity series for individual cell trajectories, and the final results are called wavelet power spectrum (WPS). WPS is typically displayed in a 2D plane, with time on the horizontal axis and frequency on the vertical axis, which vividly indicates how spectral values change over time and frequency. Thus, one is able to obtain more insights into cell migration based on the WPS rather than the Fourier power spectrum (FPS), which is only applicable to stable time series and just shows correlations with frequency.

To clarify the time-dependent properties of the WPS, we plot the WPS at each time point (also referred to as local WPS) in different *log-log* figures (Fig. 1B). Due to the extremely short sampling time, the local WPS can be viewed as the results of one stable time series. According to the published article [33], it's reasonable to fit the local WPS with the theoretical power spectrum of OU-process that is commonly used to fit FPS (Fig. 1C). After performing fits along the time axis of WPS, we definitely obtain three parameters at each time point, *i.e.*, persistence time $P_t$, migration speed $S_t$ and positional error $\sigma_t$. The first two parameters are widely used to estimate a cell's ability to move. Thus, both are named after an exclusive noun, "motility parameter".

We next introduce a classical cell motility model, termed "persistent random walk" (PRW), to simulate isotropic 2D cell migration trajectory with the fitted motility parameters at each time point as input. Therefore, we naturally obtain a set of trajectories that correspond one-by-one to all migration velocities (or time points) for individual cell trajectories (Fig. 1D). For convenience, we use an abbreviation "TSCT" (trajectories based on single-cell trajectory) to refer to the set of trajectories simulated by the PRW model with the fitted motility parameters as input from single-cell trajectory, and the corresponding process is called "single-cell trajectory reconstruction" (SCTR). So far, one can actually follow the procedures above to produce many trajectories based on a single trajectory and further study the time-dependent characteristics of migration trajectories by computing physical measures, such as MSD, VAC, and FPS, of the TSCT at each time point.

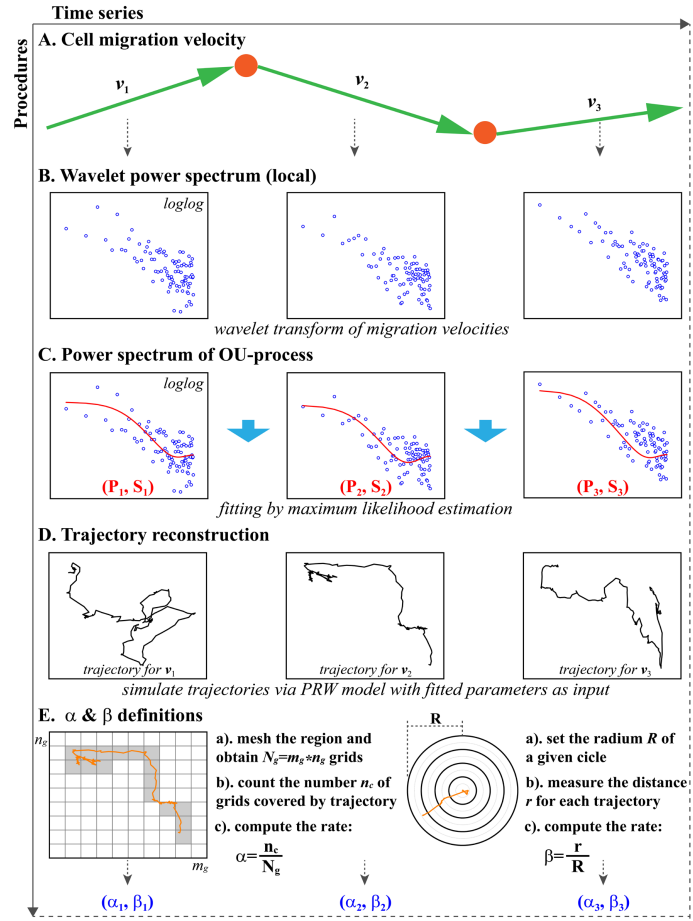

**Fig. 1** Real-time trajectory reconstruction algorithm for single-cell migration trajectory. (A) Schematic diagram of a 2D cell migration trajectory consisting of position coordinates (red dots) and migration velocities (green arrows). The velocities are obtained from the net displacements divided by the sampling time $\Delta t$. (B) Wavelet power spectrum of the migration velocity series, plotted on a *log-log* axis. (C) Motility parameters are derived by performing fits of the wavelet power spectrum at each time point with the theoretical power spectrum of the OU-process. The red lines represent the consequences of the fitting, which are dominated by persistence time $P$, migration speed $S$, and positional error $\sigma_p$. (D) Individual migration trajectories are simulated by running a persistent random walk model with the fitted parameters as input. (E) Target-finding metric $\alpha$ and migration metric $\beta$ are defined to quantify the dynamic characteristics of migration trajectories. The former is computed by $\alpha = n_c/N_g$, and the latter is computed by $\beta = r/R$. Here, the $N_g$ is the number of grids in a meshed region, the $n_c$ is the number of grids

covered by a trajectory, the $r$ is the net displacement of a trajectory and the $R$ is the radius of a predefined circle.

**Target-finding and migration metrics.**
To obtain more uniquely insights from individual cell trajectories, we define two dynamic metrics by combining our previous works [29,31], *i.e.*, target-finding metric $\alpha$ and migration metric $\beta$ (Fig. 1E). For the $\alpha$ metric, we first preset a region with a definite size of width and length, and then mesh the region to obtain $N_g = n_g * m_g$ grids. Subsequently, plot TSCT on this region and count the number $n_c$ of grids covered by individual trajectories. Finally, the ratio $n_c/N_g$ is defined as $\alpha$ (>0). Here, the unique condition defining the region is that all trajectories can be plotted and don't extend beyond the boundaries of this region.

Similarly, we first preset a circle of radius $R$ and then get the distance $r$ by directly calculating the Euclidean distance between the start and end points of each trajectory in TSCT. The $\beta$ is defined as the ratio $r/R$ (>0). Here, the radius $R$ is simply a scale to measure the length of the trajectories. Therefore, there is no limit to the value of $R$, but it is greater than 0. In this work, the averaged distance of trajectories is recommended to assign $R$. Although some concepts have been introduced in previous publications to describe target-finding and migration behaviors, they have not been combined to explore the time-varying and statistical properties of migration dynamics in terms of the trajectories of individual cells or cell populations.

**Phase diagrams of dynamic metrics indicate an optimal domain of motility parameters.**
In the foregoing sections, we have proposed a trajectory reconstruction approach and defined two dynamic metrics. To elucidate the superior characteristics of the approach, we next utilize the PRW model to simulate cell migration trajectories regulated by extracellular or intracellular cues with isotropic and constant properties. In computer simulations, the persistence time $P$ increases linearly from 0.2 to 20 min with an increment of 0.2 min, as well as migration speed $S$ from 0.1 to 10 $\mu$m/min with an increment of 0.1 $\mu$m/min, thus producing 10,000 combinations of parameters $P$ and $S$. For each parameter combination, we repeatedly run the PRW model 200 times, and consequently obtain 200 sets of metrics $\alpha$ and $\beta$ and their respective averages.

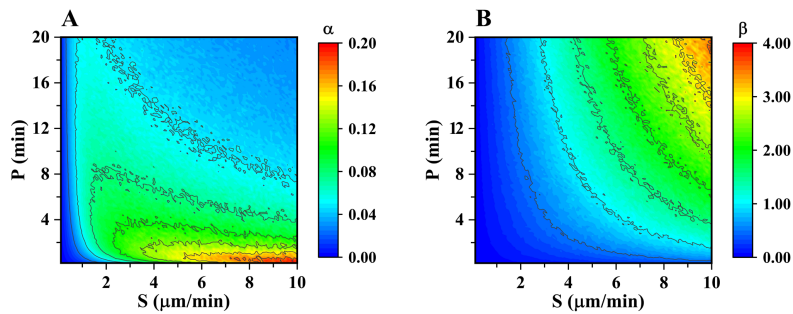

**Fig. 2** Phase diagrams of target-finding metric $\alpha$ and migration metric $\beta$ obtained from PRW simulations. (A) Phase diagram of $\alpha$ regulated by persistence time $P$ that increases from 0.2 to 20 min with an increment of 0.2 min and migration speed $S$ from 0.1 to 10 $\mu$m/min with an increment of 0.1 $\mu$m/min. (B) Phase diagram of $\beta$ regulated by the same parameter sets in (A). The color bars indicate the amplitudes of the two metrics, while the solid lines represent the contours. Note that each value in the two diagrams denotes the average of 200 metrics $\alpha$ and $\beta$, respectively.

In Fig. 2, the phase diagrams further clearly present the correlations of averaged $\alpha$ and $\beta$ with $P$ and $S$, respectively. For the averaged $\alpha$ (Fig. 2A), the diagram first shows that the $\alpha$-peak (colored in red) appears in the lower-right region, where the parameter $S$ is roughly in the range of 8-10 $\mu$m/min and the $P$ is in the range of 0-1 min (see the thin contour lines). This means that a cell with motility parameters within this range will cover more grids or search more areas at a given time. In addition, the results also indicate that the $\alpha$ is strongly correlated with the parameters $P$ and $S$ in diverse manners, including stable, non-monotonic, and monotonic manners. For instance, in the interval of $S<0.5$ $\mu$m/min, the $\alpha$ is close to 0 and almost does not change with the increase of $P$, which can be viewed as a stable manner. By contrast, the $\alpha$ first increases gradually and then decreases in the interval of $0.5<S<7.5$ $\mu$m/min, and decreases continuously in the interval of $S>7.5$ $\mu$m/min, implying that the stable manner becomes non-/monotonic manners. Similarly, the $\alpha$ also undergoes a "first increases-then decreases" process with the increase of $S$ in the interval of $P>2$ min, and increases persistently in the interval of $P<2$ min. In contrast, the phase diagram in Fig. 2B indicates that the $\beta$-peak appears in the upper-right region, meaning that the $\beta$ monotonously increases as the parameters $P$ or $S$ increase.

Taken together, there are four key aspects that deserve further attentions: i) the metric $\alpha$ is strictly regulated by the combination of $P$ and $S$, ii) the parameter $S$ is seemingly more important than the $P$ because the $\alpha$-peak is closer to the horizontal ($S$) axis, iii) there is an optimal domain of $P$ and $S$ that corresponds to a maximal value of $\alpha$, and iv) the $\beta$ possesses remarkably different correlations with motility parameters from the $\alpha$.

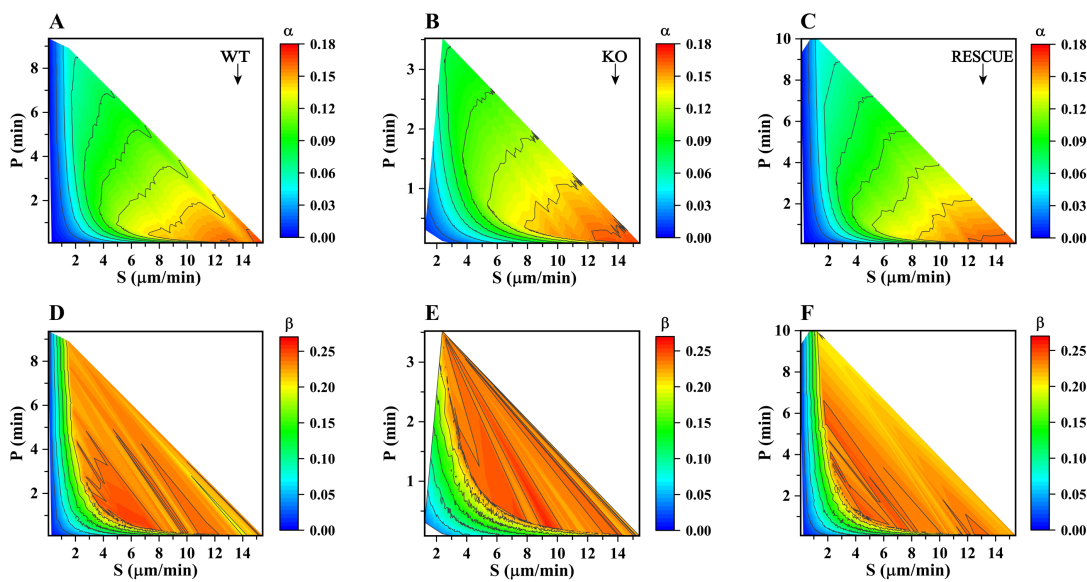

**Fig. 3** Phase diagrams of target-finding metric $\alpha$ and migration metric $\beta$ obtained from experimental data of *D. discoideum* (also known as the *social amoeba*) migration. (A) Phase diagram of $\alpha$ for the wild-type (WT) amoeba group. (B, C) Phase diagrams of $\alpha$ for the Arpin knocked-out (KO) and rescued (RESCUE) amoeba groups. (D, E, and F) Phase diagrams of $\beta$ for the same WT, KO, and RESCUE amoeba groups, respectively.

To validate the reconstruction approach and the novel results shown in Fig. 2, we further compute the time-varying dynamic metrics $\alpha$ and $\beta$ of each trajectory in *D. discoideum* migration data. See Fig. S1 in the *Supplementary Material* for more details on the trajectories. We found that the changing trends in Fig. 2 also appear in Fig. 3. Firstly, the phase diagrams of $\alpha$ and $\beta$ possess some similarities with those in Fig. 2A, including i) single $\alpha$-peak appears in each diagram for the WT, KO and RESCUE groups, ii) all of the $\alpha$-peaks locate in the lower-right region and are closer to the horizontal axis (Fig. 3A-C), and iii) single $\beta$-peak also appears in the diagrams (Fig. 3D-F) with the directions toward upper-right region. It should be noted that the blank regions in the diagrams are caused by the fact that some combinations of the fitted motility parameters are not obtained due to the limited experimental data. Although the experimental results shown in Fig. 3 are quantitatively different from those in Fig. 2, there is a high degree of consistency at the qualitative level, which further validates the superior performance of the reconstruction approach, as well as the existence of optimal domains of motility parameters experimentally. In the *Supplementary Material*, we also thoroughly discuss the characteristics of subregion in the diagrams of the metrics and further analyze the self-similar properties encoded in the phase diagrams (Fig. S2), by which the results in Fig. 2-3 could prove to be consistent qualitatively.

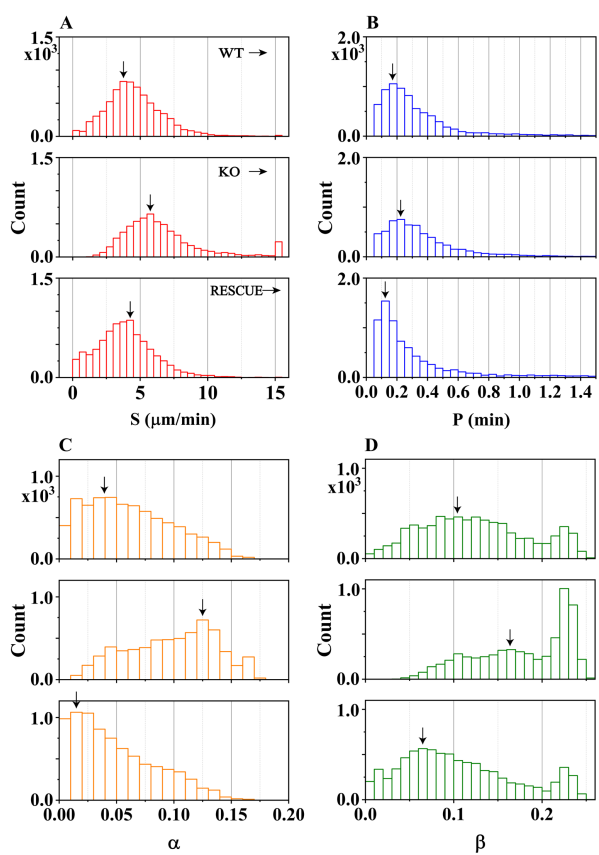

**Fig. 4** Statistical analysis of the time-dependent motility parameters and the dynamic metrics. (A) Statistical histograms of migration speed $S$ for the WT (top), KO (middle), and RESCUE (bottom) amoeba groups. (B) Statistical histograms of persistence time $P$. (C) Statistical histograms of target-finding metric $\alpha$. (D) Statistical histograms of migration metric $\beta$. The black arrows mark the peaks of individual histograms.

**Statistical measures of migration potential regulated by Arpin protein**

To explore more information from the derived dynamic metrics $\alpha$ and $\beta$, we next focus on statistical measures to evaluate the migration potential of three groups of *D. discoideum* (*i.e.*, WT, KO, and RESCUE groups). Firstly, we gather all the time-varying dynamic metrics and motility parameters for all migration velocities in each group, and subsequently draw the respective statistical histograms, as shown in Fig. 4. In contrast, the results in Fig. 4A show that the overall $S$, 5.74 $\mu$m/min (marked by the black arrow) of the KO group is remarkably greater than 4.05 $\mu$m/min of the WT and 4.26 $\mu$m/min of the RESCUE groups. And another persistence time $P$ also accords with the relative relationship above, *i.e.*, the overall $P$ 0.22 min of the KO group is greater than 0.17 min and 0.12 min of the WT and RESCUE groups, respectively (Fig. 4B).

In addition, the results in Fig. 4C show that the histogram of the KO group possesses a significantly different shape from the other two groups, and the corresponding overall $\alpha$ 0.13 is obviously greater than 0.04 and 0.02 of the WT and RESCUE groups. Similarly, the overall $\beta$ 0.17 of the KO group is also greater than 0.10 and 0.07 of the WT and KO groups (Fig. 4D). Certainly, it's more vivid that the peak of the RESCUE group is closest to the y axis, followed by the WT and KO groups, for the dynamic metrics. On the whole, the migration potential of the KO group is greater than that of the other two groups, suggesting that the loss of Arpin protein improves the ability of target-finding and migration in *D. discoideum*. More importantly, although both the WT and RESCUE amoeba are control groups and significantly different from the KO group, the RESCUE group intrinsically possesses different dynamic properties from the WT group, which has not been reported in the previous work [34].

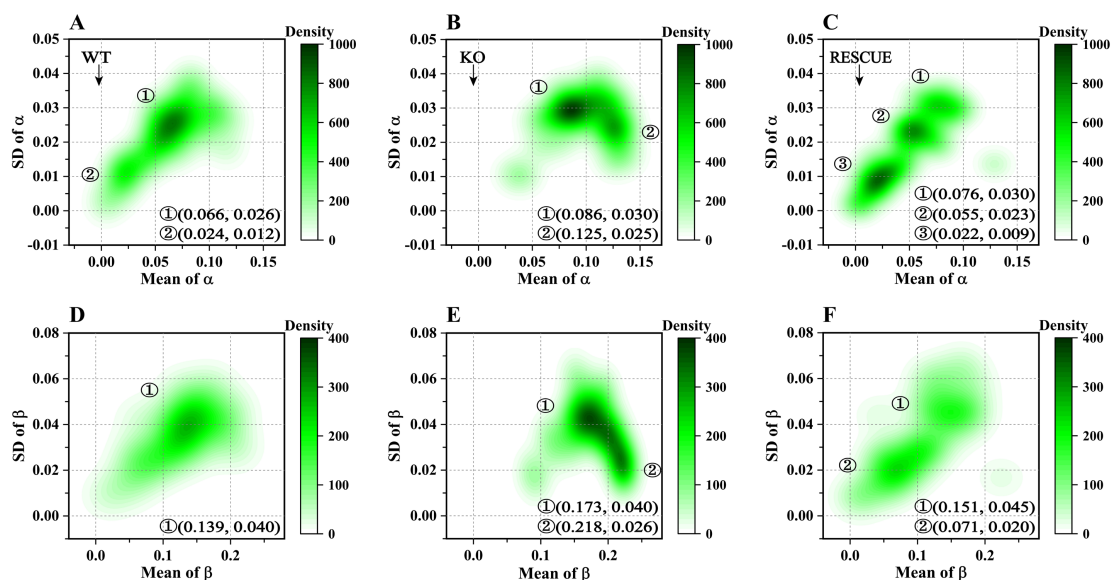

**Fig. 5** Population dynamics of *D. discoideum* migration. (A) Population characteristics are represented together by the mean and SD (standard deviation) of the time-varying $\alpha$ for individual trajectories in the WT group. (B, C) Population characteristics for the KO and RESCUE groups, respectively. (D, E, and F) Population characteristics represented by the mean and SD of the time-varying $\beta$ for individual trajectories in the three groups. Note that the numbers ①~③ are used to indicate the peaks of the 2D distributions for the three groups.

In order to further clarify the differences between the three groups of *D. discoideum*, we again focus on the dynamic metrics $\alpha$ and $\beta$ and analyze their statistical variations from a novel perspective. Before conducting the statistical analysis, we first calculate the mean and SD of the time-varying dynamic metrics for the single-cell migration trajectories in each group. Subsequently, a 2D density plot of scatter is created based on kernel density estimation, which actually shows the population dynamics of each group by setting the mean on the x-axis and the SD on the y-axis, as shown in Fig. 5. The results firstly illustrate that the density plots of all groups contain several peaks colored in black regardless of the metrics, and the peaks correspond to the main population characteristics. In other words, one can directly evaluate the main migration modes in a population according to the distributions of the peaks. For the metric $\alpha$, we observe that there are two peaks in the density plot of the WT group (Fig. 5A), one is conspicuous at the position of (0.066, 0.026) and another is inconspicuous at (0.024, 0.012), which are similar respectively to the second (0.055, 0.023) and third (0.022, 0.009) peaks of the RESCUE group, apart from the first peak (0.076, 0.030) (Fig. 5C). Although these two plots are similar to each other, neither of them is similar to the plot of the KO group that contains two peaks at the positions of (0.086, 0.030) and (0.125, 0.025) (Fig. 5B).

In contrast to the metric $\beta$, the density plot of the WT group only contains a peak at the position of (0.139, 0.040) (Fig. 5D), and it differs from the plot of the RESCUE group with two subtle peaks at (0.151, 0.045) and (0.071, 0.020) (Fig. 5F). Moreover, the density plot of the KO group also shows significant characteristics, *i.e.*, both of the two peaks are remarkable at (0.173, 0.040) and (0.218, 0.026), and are farther away from the y axis relatively (Fig. 5E).

According to the results in Fig. 5, we can infer more information about the three groups of *D. discoideum* regulated by Arpin protein, *i.e.*, i) how the population dynamics is distributed for any group, ii) the migration potentials of the WT and RESCUE groups are significantly different from that of the KO group, and iii) the potential of the RESCUE group is also finely distinguishable from that of the WT group. The analysis above enhances the results in Fig. 4 in a straightforward way, allowing us to further discuss some interesting issues, such as the reason for the appearance of three peaks for the $\alpha$ of the RESCUE group and the cause of the visible differences between the RESCUE and WT groups.

**Time-varying metrics reveal real-time characteristics of population dynamics.**
In this section, we continue to analyze the dynamic metrics for revealing the time-dependent migration characteristics. Firstly, we sequentially stack the time-varying $\alpha$ series of single-cell trajectories along the vertical axis and subsequently obtain a 2D heatmap that correlates with elapsed time (x-axis) and cell No. (y-axis), as seen in Fig. 6. In comparison, we found qualitatively that there are more $\alpha$ values close to the maximum 0.18 (in red) in the KO group (Fig. 6B), while there are more $\alpha$ values close to the minimum 0 (in cyan) in the RESCUE group (Fig. 6C), and it seems that the WT group is between the KO and RESCUE groups (Fig. 6A). The qualitative descriptions above directly illustrate that a larger $\alpha$ dominates in the KO group while a smaller value does in the RESCUE group.

Further, the ensemble-averaged $\alpha$ series are computed by averaging all $\alpha$ values along the vertical (cell No.) axis. The results clearly indicate that the ensemble-averaged $\alpha$ of the WT group are subtly greater than those of the RESCUE group at most time points, while all $\alpha$ values of the two groups above are significantly less than the values of the KO group and a horizontal line of $\alpha=0.75$ can be used to distinguish the significant difference (Fig. 6D). And the three ensemble-averaged series also vividly show how the $\alpha$ values evolve with time, such as the emerging peak around 10.4 (*i.e.*, 125*5/60) min for the KO group. In addition, we further obtain the temporal-averaged $\alpha$ values by averaging all $\alpha$ along the horizontal (time) axis, and the results show that there are significant fluctuations between cells, illustrating that individual differences are apparent. Subsequently, the ensemble-temporal-averaged $\alpha$ values, *i.e.*, 0.061±0.005 for the WT, 0.098±0.004 for the KO, and 0.049±0.004 for the RESCUE groups in the black rectangle, are significantly different from each other (Fig. 6E). These differences are also consistent with the results in Figs. 4 and 5.

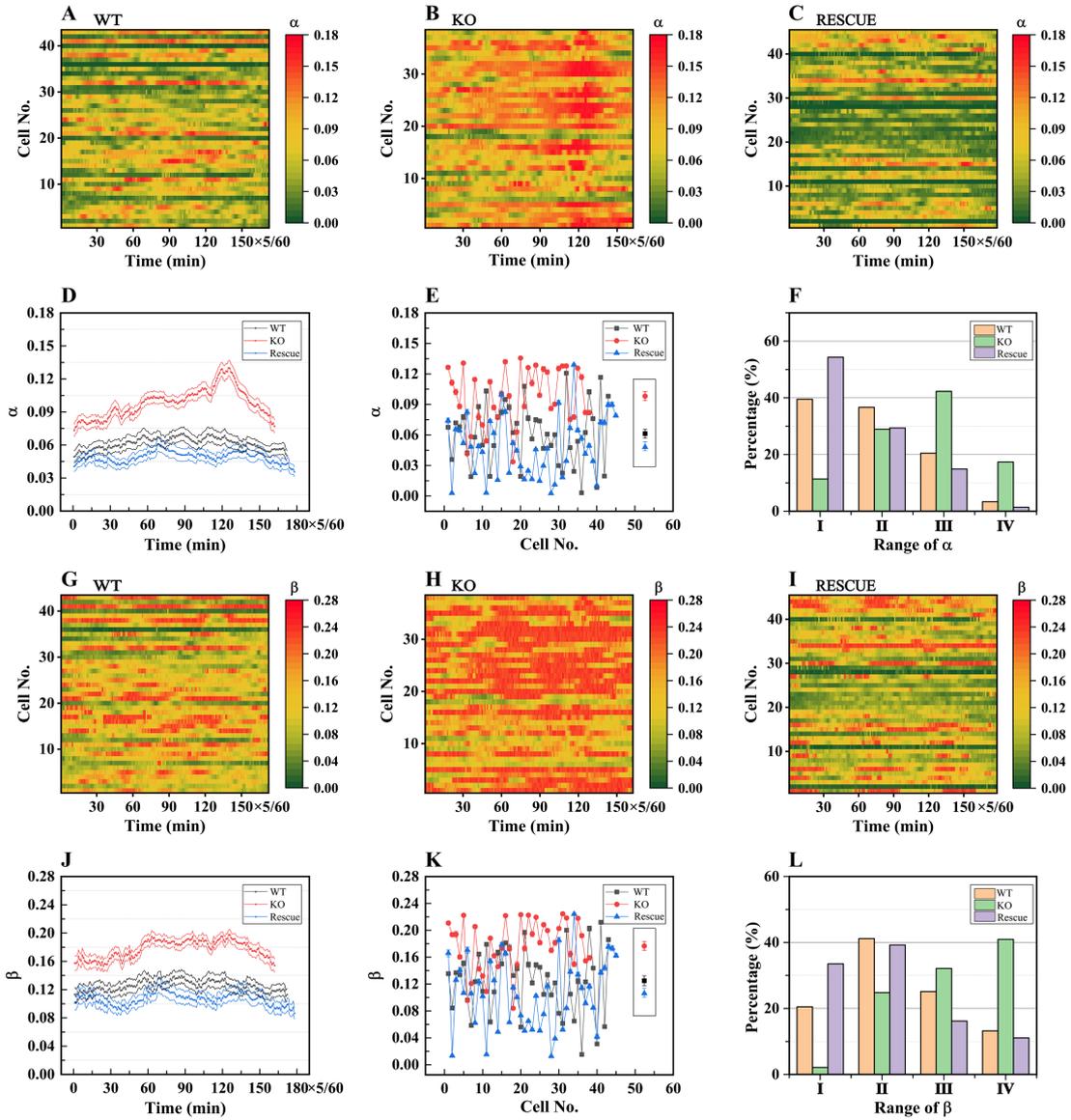

**Fig. 6** Real-time characteristics of population dynamics measured by the time-varying target-finding and migration metrics. (A, B, and C) Target-finding metric $\alpha$ as a function of elapsed time, respectively for the WT, KO, and RESCUE groups. The heatmaps reveal the relationships between the metrics, time, and cell number, which are obtained by sequentially stacking each $\alpha$ profile along the vertical axis. (D) Ensemble-averaged $\alpha$ profiles (thick lines in black, red, and blue) against elapsed time for the three groups. The thin lines next to the thick lines denote SEM (standard error of mean) for 43, 38, and 45 cells. (E) Temporal-averaged $\alpha$ values against cell number for the three groups. The symbols in the black rectangle denote the mean±SEM of the temporal-averaged $\alpha$ values, *i.e.*, ensemble-temporal-averaged values. (F) Percentages of the $\alpha$ in four different ranges, based on the data in (A-C). (G, H, and I) Same captions as those in (A-C) but for migration metric $\beta$. (J, K, and L) Same captions as those in (D-F) but for migration metric.

Next, we divide the whole range (0-0.18, from cyan to red) of the $\alpha$ at all time points into four intervals (*i.e.*, I: 0-0.045, II: 0.045-0.09, III: 0.09-0.135, IV: 0.135-0.18), and then count the number and compute the percentage of the $\alpha$ in each interval (Fig. 6F). The corresponding results indicate that the percentage of the WT group follows a gradually decreasing trend from 39.52% to 3.35%, which is highly similar to the trend from 54.32% to 1.4% of the RESCUE group. By contrast, the percentage of the KO group first increases and then decreases, and reaches a maximum of 42.3% in interval III. These changes are consistent with the results in Fig. 4C and further improve the intelligibility of the results to some extent.

Following the same procedures used in the analysis of the time-varying $\alpha$, we continue to study the time-dependent characteristics of the $\beta$. The heatmaps (Fig. 6G, H, and I) present qualitatively similar results to those of the $\alpha$ in Fig. 6A-C, *i.e.*, a larger $\beta$ close to 0.28 (in red) dominates in the KO group while a smaller $\beta$ close to 0 (in cyan) does in the RESCUE group. Similarly, the ensemble-averaged $\beta$ values of the KO group are significantly greater than those of the WT and RESCUE groups, and the values of the WT group are generally greater than those of the RESCUE groups (Fig. 6J). Furthermore, the ensemble-temporal-averaged $\beta$, *i.e.*, 0.125±0.0077, 0.177±0.006 and 0.106±0.008 for the WT, KO, and RESCUE groups, show a similar relationship to that in Fig. 6E (Fig. 6K). Finally, the percentages of the WT and RESCUE groups first increase and then decrease, reaching maximums of 41.18% and 39.24%, respectively, in interval II. However, the percentage of the KO group progressively increases from 2.14% to 40.95% (Fig. 6L). Here, the changing trend of the percentage is the only difference from that of the $\alpha$ in Fig. 6F.

**Emergent periodic behaviors of metrics are strongly correlated with migration potential**

On the basis of the time-varying dynamic metrics $\alpha$ and $\beta$, we continue to investigate how the two metrics evolve with elapsed time for individual trajectories. When the $\alpha$ and $\beta$ series of single-cell trajectory are plotted in 2D plane, we found that some curves change regularly with time, as seen in Fig. 7A. Both of the $\alpha$ and $\beta$ series change periodically from a considerable value to a small value (or inversely) in a constant time interval, and behave like a *sine*, *cosine* or other function. To explain the periodic behaviors, we further compute the correlation coefficients between the time-varying metrics and the motility parameters for each trajectory. The results in Fig. 7B indicate that all of the correlations ($r_{p\alpha}$) of the $\alpha$ with the $P$ are less than 0.20, and in particular, the $r_{p\alpha}$ of the KO group is negative (-0.08±0.07). In contrast, all the correlations ($r_{s\alpha}$) of the $\alpha$ with the $S$ are greater than 0.44. These two contrary features have been proven to be statistically significant (***$p<0.001$). It should be noted that the motility parameters are normalized in advance to the closed range of 0-1 to eliminate the magnitude difference between the parameters and the metrics, when calculating the correlation coefficients.

However, for the time-varying $\beta$, all of the correlations ($r_{p\beta}$ and $r_{s\beta}$) of the $\beta$ with the $P$ and $S$ are greater than 0.25, and the $r_{s\beta}$ of the WT and KO groups are slightly greater than the $r_{p\beta}$ of the two groups, while there is no significant difference ($p>0.05$) between the two coefficients for the RESCUE group (Fig. 7C). Therefore, it can be inferred that the time-varying $\alpha$ and $\beta$ are consequences of the combinations of the parameters $P$ and $S$, and the $S$ imposes a relatively more significant effect on the $\alpha$, while the two parameters exert almost the same effect on the $\beta$. See Fig. S3 for further discussion of the roles of the parameters in regulating the metrics. It can be concluded that although the $P$ and $S$ change irregularly with time, the dynamic metrics emerge periodic behaviors under the combined influence of these two parameters. For a more vivid contrast, see Fig. S4 in the *Supplementary Material*.

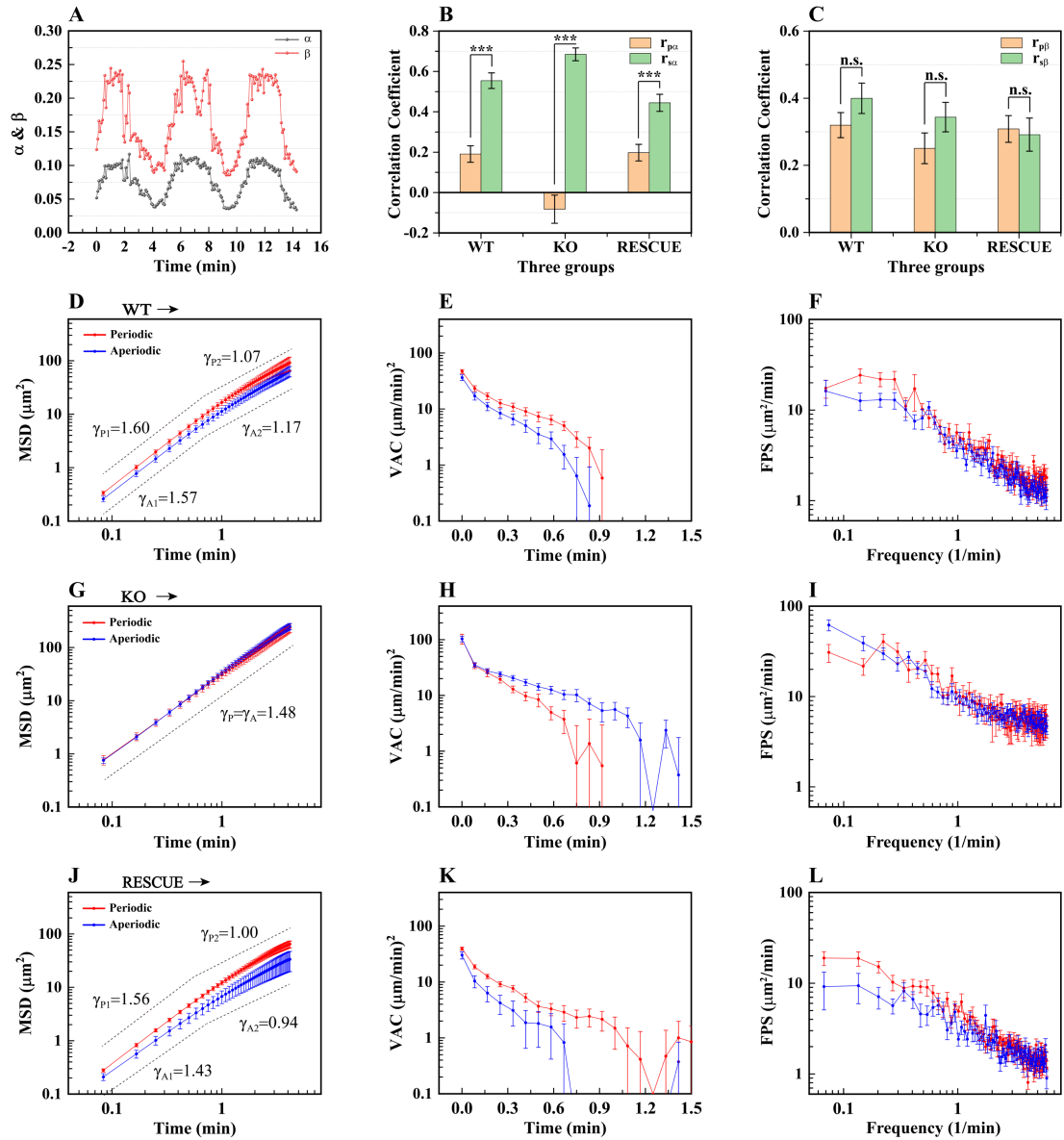

**Fig. 7** Emergent periodic behaviors of $\alpha$ and $\beta$ profiles are correlated with the migration potential. (A) Representative $\alpha$ (black) and $\beta$ (red) profiles change periodically with elapsed time. (B) The correlations of the $\alpha$ with persistence time $P$ ($r_{p\alpha}$ in orange) and migration speed $S$ ($r_{s\alpha}$ in green) are represented by histograms. Data are mean±SEM (standard error of the sample mean); the cell numbers are 43, 38, and 45, respectively; ***$p<0.001$, Wilcoxon rank sum test for the WT and KO groups, $t$-test for the RESCUE group. (C) The correlations of the $\beta$ with persistence time $P$ ($r_{P\beta}$ in orange) and migration speed $S$ ($r_{s\beta}$ in green). Data are presented as mean±SEM; the cell numbers are 43, 38, and 45, respectively; *n.s.* denotes non-significance, $t$-test for the three groups. (D) MSDs of *D. discoideum* migration for periodic (in red) and aperiodic (in blue) $\alpha$ profiles in the WT group. (E) VACs for periodic and aperiodic $\alpha$ profiles. (F) FPS also for the two behaviors. (G, H, and I) Same captions as those in (D-F), but for KO group. (J, K, and L) Also, the same captions as those in (D-F), but for RESCUE group. The data in (D-L) are denoted by mean ±SEM.

In order to make certain whether the periodic behavior is correlated with migration potential,

we further artificially classify all $\alpha$ series into periodic and aperiodic categories for each group. It should be noted that the averaged correlation coefficient $r_{\alpha\beta}$ is 0.88±0.10 for the three groups of *D. discoideum*. Therefore, there are few significant differences in which metric is chosen as the classification object. Subsequently, we compute three classical physical quantities, namely MSD, VAC, and FPS, to measure accurately the migration potential of the two categories for the three groups of *D. discoideum*. The results show that the MSD profile of the periodic WT group first increases rapidly with a slope of $\gamma_{P1}$=1.60 and then increases slowly with a slope of $\gamma_{P2}$=1.07, and the MSDs are almost greater than those of the aperiodic WT group that are characterized by the slopes of $\gamma_{A1}$=1.57 and $\gamma_{A2}$=1.17 (Fig. 7D). However, the MSD profiles of the periodic and aperiodic KO groups gradually increase with an identical slope of 1.48, and there is no significant difference between the two MSD profiles (Fig. 7G). After being rescued by the Arpin protein, the MSD profile of the periodic RESCUE group also sustainably grows with two slopes of 1.56 and 1.0, and the corresponding MSDs are significantly greater than those of the aperiodic RESCUE group dominated by two different slopes of 1.43 and 0.94 (Fig. 7J).

In addition, the VAC profiles of the WT group show similar changing trends to those of the RESCUE group, *i.e.*, the VACs of the periodic group are nearly greater than those of the aperiodic group for the WT and RESCUE groups, and all of the VAC profiles follow a nonlinear decay process in the entire time interval (Fig. 7E and K). In contrast, the VAC profiles of the KO group exhibit opposite trends, namely, the VACs of the aperiodic KO group are greater, excluding the first three values, and the VAC profiles follow a linear decay process, except for the first value (Fig. 7H). Moreover, the results in VAC profiles are further enhanced by the FPS profiles in Fig. 7F, I, and L.

Taken together, the emergent periodic behaviors of dynamic metrics are strongly correlated with the migration potential of *D. discoideum*; specifically, the potential is stronger in periodic WT and RESCUE groups. In other words, periodical behavior corresponds to a stronger migration potential. For the KO group, there is no significant difference in the MSD profiles; however, a significant difference is observed in the VAC profiles. We speculate it's the loss of Arpin protein that disrupts the high correlations.

**Discussion**

In this paper, we proposed a single-cell trajectory reconstruction approach, which mainly combines the calculations of wavelet power spectrum of migration velocities, the fits of the power spectrum of an OU-process for deriving motility parameters, trajectory simulations based on PRW model, and the analysis of dynamic metrics $\alpha$ and $\beta$ (Fig. 1). The approach allows us to estimate the migration potential of cells based on individual trajectories, including statistical and time-varying properties, and further reveal how the potential is affected by different motility parameters, intracellular crucial proteins and distinctive migration modes.

With the reconstruction approach, we first investigate the quantitative relationships between motility parameters ($P$ and $S$) and dynamic metrics ($\alpha$ and $\beta$) by running the PRW model to simulate cell migration trajectories on a computer. In general, the two dynamic metrics show significantly different changing modes, *i.e.*, the $\alpha$ is correlated nonmonotonically with the motility parameters (Fig. 2A). In contrast, the $\beta$ is correlated monotonically with the parameters (Fig. 2B). In other words, the $\alpha$ reaches maximum when the parameters are in a constant range, *i.e.*, 0<$P$<1 min and 7.5<$S$<10 $\mu$m/min, and the $\beta$ reaches maximum when both of the parameters are larger. In particular, the different modes also indicate that the $\alpha$ is intrinsically different from the $\beta$. Since the $\alpha$-peak is closer to the horizontal ($S$) axis and is shaped like a strip, the $\alpha$ is relatively more sensitive to the $S$, or the $S$ plays a more critical role in regulating the $\alpha$. This claim is reasonable and can be verified by two limiting cases: the $\alpha$ must be zero when the $S$ is zero, whereas the $\alpha$ is hardly zero when the $P$ is zero. More interestingly, for a large value of $P$, cell migration behaves like ballistic motility, and the cell will pass through a target region in a directed manner but is likely to miss the target. By contrast, for a large value of $S$, cell migration behaves like diffusive motility, and the cell will search a target region in a chaotic and random way. Following the procedures used to analyze simulated trajectories, we further analyze three groups of *D. discoideum* data (Fig. 3 and Fig. S1), and the results are qualitatively consistent with the changing modes shown in Fig. 2 (Fig. S2). Based on the analysis in the paper, we will further optimize the approaches and models to reproduce novel results that are quantitatively identical to those of experimental data in the follow-up research.

In addition to the changing modes encoded in phase diagrams, we also reveal the differences between the three groups of *D. discoideum* migration data from a statistical perspective. The results firstly indicate that the $P$ values of the KO group are larger than those of the WT and RESCUE groups, and the distribution of the WT group is similar to that of the RESCUE group. Likewise, the $S$ values of the three groups essentially follow the relationships above; however, in detail, the values of the RESCUE group are closer to zero than those of the WT group (Fig. 4A, B). Remarkably, the distributions of the two dynamic metrics enlarge the differences and further verify the relationships mentioned (Fig. 4C, D). To obtain more insightful information, we further analyze the population characteristics of the three groups based on the two metrics. The density plots of "*mean-SD*" vividly show how the time-dependent metrics are distributed for each group, such as the peaks or clusters, and how different the population characteristics are from the other groups (Fig. 5). As a whole, the analysis above illustrates the differences between the WT, KO and RESCUE groups in terms of statistics, and especially the unreported result that the RESCUE group is not identical to the WT group on the detailed level, which we believe is mainly caused by the content of Arpin proteins and/or the experimental procedures.

Along with the statistical characteristics, we further investigate the time-varying properties of the two dynamic metrics. The results in heatmaps not only clearly show how the two metrics evolve over time for single-cell trajectory but also reveal how the population properties change with time for each group (Fig. 6A-C, G-I). In particular, one can directly estimate or compare the values of metrics for different cells or groups according to the color bars; for instance, there are more values in red in the KO group and more in cyan in the RESCUE group for the $\alpha$ and $\beta$. Therefore, it's reasonable to deduce that the abilities of target-finding and migration are stronger in the KO group than the other two groups. Moreover, the ensemble-averaged metrics also exhibit the time-dependent properties of each group, such as first increasing and then decreasing, forming an obvious peak at a specific time interval, and fluctuating persistently over time. Subsequently, the ensemble-temporal-averaged values further highlight the differences between the groups and enhance the result that the RESCUE group is indeed distinct from the WT group (Fig. 6D-F, J-L).

Finally, we focused on the metric profiles of single-cell trajectories and found there are periodic behaviors that change regularly over elapsed time (Fig. S4). To explain the behaviors, we calculate correlations of dynamic metrics with motility parameters, respectively. The results show that the parameter $P$ is weakly correlated with the $\alpha$ (<0.2), whereas the $S$ is highly correlated with the $\alpha$ (>0.44), meaning that the $S$ dominates in regulating the $\alpha$. Furthermore, the $P$ exerts a similar effect on the $\alpha$ (~0.2) in the WT and RESCUE groups, and the $P$ is negatively correlated with the $\alpha$. Differently, the impact of the $P$ is largest in the KO group, followed by the WT and RESCUE groups. In contrast, both the motility parameters are positively correlated with the $\beta$, and the $S$ exerts a subtly larger effect on the $\beta$. Therefore, the motility parameters play a crucial role in determining the periodic behaviors, and their impact varies across each group (Fig. 7A-C and Fig. S3). Next, we artificially classify each group into the periodic and aperiodic groups and compare the migration dynamics of each subgroup using MSD, VAC, and FPS. The results show that for the WT and RESCUE groups, the MSDs of the periodic group are significantly larger than those of the aperiodic group. In contrast, there is no significant difference between the MSDs of the two subgroups for the KO group (Fig. 7D, G, J). Furthermore, the VACs also follow the same relationship for the WT and RESCUE groups, but an opposite relationship for the KO group (Fig. 7E, H, K). Likewise, the FPSs in the low-frequency domain (<0.4 /min) also validate the results shown by the VACs (Fig. 7F, I, L). As a whole, the analysis above illustrates that the motility parameters determine the emergent periodic behavior, and it is also correlated with migration potential.

**Conclusion**

In this paper, we propose an approach to reconstruct single-cell trajectories, which primarily combines a wavelet transform, fits of power spectrum, trajectory simulation using a dynamic model, and calculations of dynamic metrics, enabling us to investigate cellular potentials in terms of target-finding and migration. Our analysis has revealed diverse and complex relationships between the motility parameters and dynamic metrics, particularly the existence of an optimal parameter range, and clarified the role of each parameter in regulating the metrics. Moreover, the approach also reveals the statistical properties of *D. discoideum* migration and further clarifies population differences between the three treated and control groups. Notably, the loss of Arpin protein improves the abilities of target-finding and migration, and the previously unreported result is that the RESCUE group actually differs from the WT group. Additionally, we systematically analyzed the time-dependent properties of *D. discoideum* migration and found that periodic behaviors of the two metrics can emerge naturally under the combined effect of motility parameters that change irregularly over time, which strongly correlates with the migration potential. Overall, the approach enables the precise analysis of how the characteristics are distributed and how they evolve over time, and it further elucidates the effects of intracellular Arpin protein on the modes and behaviors of *D. discoideum* migration.

**Methods**

***D. discoideum* migration data**

In this work, we analyzed the migration data of Dictyostelium discoideum (*D. discoideum*) amoeba, which contains wild-type (WT) amoeba, Arpin-knockout (KO) amoeba, and rescued (RESCUE) amoeba by green fluorescent protein (GFP)-Arpin expression in KO amoeba. The experimental data here were obtained with permission from the published work by Gautreau *et al*. in *Nature* journal. See works [29,31,34] for more details on the experiments and analysis.

**PRW model**

In this paper, we mainly utilize the PRW model to perform two tasks: the first is to simulate cell migration trajectories with the prescribed motility parameters (persistence time $P$ and migration speed $S$) as input for exploring the relationships between these parameters and dynamic metrics (target-finding $\alpha$ and migration $\beta$); the second is also to simulate trajectories but based on the motility parameters fitted from the WPS of individual migration velocity series. Specifically, the position coordinates of one cell migration on an isotropic 2D plane can be updated iteratively by the following formula:

$$\boldsymbol{r}(t + \Delta t) = \boldsymbol{r}(t) + \Delta \boldsymbol{r}(t, \Delta t), \tag{1}$$

where the $\boldsymbol{r}$ is the position vector that can be represented by two components of $x$ and $y$, $\Delta \boldsymbol{r}$ is the displacement between any two successive positions, and $\Delta t$ is the time step that corresponds to the experimental sampling time $\delta t$.

Further, the $\Delta \boldsymbol{r}$ can be computed using the formula:

$$\Delta \boldsymbol{r}(t, \Delta t) = \kappa \cdot \Delta \boldsymbol{r}(t - \Delta t, \Delta t) + \xi \cdot \boldsymbol{w}, \tag{2}$$

where the $\kappa$ is written as $1 - \Delta t/P$ that reflects the correlation of the second $\Delta \boldsymbol{r}(t, \Delta t)$ with the first $\Delta \boldsymbol{r}(t - \Delta t, \Delta t)$ displacements, and the $\boldsymbol{w} \sim \mathcal{N}(0,1)$ is a random Gaussian noise that is regulated by the $\xi = (S^2 \Delta t^3 / P)^{0.5}$. Here, the $P$ and $S$ are motility parameters illustrated in the previous sections, and the $\boldsymbol{r}(t)$ can be represented in the form of $\boldsymbol{r}_i$ with a subscript $i$. Before running the PRW model in computer, some arguments need to be determined first, including i) the iteration number $N_p$ (or simulation duration $T$), ii) the time step $\Delta t$, iii) the initial position $\boldsymbol{r}_1$, iv) the initial displacement $\Delta \boldsymbol{r}_1$, and v) the prescribed or fitted motility parameters. Finally, positional errors $\sigma_p \cdot \boldsymbol{w}$ are also added to the simulated trajectories to account for the effect of experimental measurements.

**MSD**

For the migration trajectory $\boldsymbol{r}_i$ of individual cells, we can calculate the velocity series $\boldsymbol{v}_i$ by the formula of $\Delta \boldsymbol{r}_i / \Delta t$ and further derive the corresponding MSD evaluation, which is commonly used to quantify the ability of cell to migrate with the following form:

$$MSD(\tau) = \frac{1}{N_p - \tau} \sum_{i=1}^{N_p - \tau} (\boldsymbol{r}_{i+\tau} - \boldsymbol{r}_i)^2, \tag{3}$$

where the $N_p$ is the number of frames (or positions) contained in a single trajectory, which is also identical to the number of iterations in computer simulations, and the $\tau$ is a variable to measure how many the time step is between any two positions.

Based on the MSD on the *log-log* axis, we not only evaluate and compare the migration ability of different cell lines but also reveal the migration modes encoded in the MSD profile. More precisely, a slope $\gamma$ of the MSD profile can be directly extracted by a formula of $log(MSD) \sim \gamma \cdot log(\tau)$ to estimate how similar the migration is to ballistic ($\gamma$=2) or Brownian ($\gamma$=1) motion, or to determine whether the migration is superdiffusion ($\gamma$>1) or subdiffusion ($\gamma$<1).

**VAC**

Different from the MSD evaluation, another quantity, VAC, is also constructed previously to characterize the correlation of one migration velocity with another with the following form:

$$VAC(\tau) = \langle \boldsymbol{v}_{i+\tau} \cdot \boldsymbol{v}_i \rangle \cong \frac{1}{N_v - \tau - 1} \sum_{j=1}^{N_v - \tau} \left( \boldsymbol{v}_j - \frac{1}{N_v - \tau} \sum_{k=1}^{N_v - \tau} \boldsymbol{v}_k \right) \left( \boldsymbol{v}_{j+\tau} - \frac{1}{N_v - \tau} \sum_{k=\tau+1}^{N_v - \tau} \boldsymbol{v}_k \right), \tag{4}$$

where the $N_v = N_p - 1$ is the number of migration velocities for individual trajectories.

According to the evaluation, one can estimate the correlation between any two velocities at a time lag of $\tau$ and directly analyze a crucial property, *i.e.*, the decay mode of the VAC profile. If the VACs decrease linearly with a time lag on the *log-lin* axis, it illustrates that the VAC profile follows a single-exponential decay function. Otherwise, the nonlinear VAC profile can be described by a superposition of several single-exponential decay functions. In general, cell migration in isotropic microenvironments assumes a linear VAC profile, whereas it becomes nonlinear in anisotropic conditions. Therefore, we can roughly determine whether migration behavior is isotropic or heterogeneous based on the analysis of the VAC profile.

**FPS**

In this work, although the two classical physical quantities, MSD and VAC, possess superior performance in the analysis of cell migration ability and directional persistence (correlations), both of them cannot avoid a pivotal issue, *i.e.*, the two quantities are highly correlated with time, and therefore they will return unreliable errors on the fitted motility parameters [35]. To avoid this issue, we introduce a third physical quantity, Fourier power spectrum (FPS) in frequency domain, which is derived by performing Fourier transform of the VAC in time domain according to Wiener-Khinchin theorem [36], *i.e.*, "the power spectrum of any generalized stationary random process is the Fourier transform of its autocovariance function". Under the guidance of the theorem, we first calculate the Fourier transform of migration velocities with the following formula:

$$\boldsymbol{v}_k = \Delta t \sum_{j=1}^{N_v} \boldsymbol{v}_j \cdot e^{i2\pi f_k t_j} = \Delta t \sum_{j=1}^{N_v} \boldsymbol{v}_j \cdot e^{i2\pi k j/N_v}, \tag{5}$$

where the $f_k$ is Fourier frequency with a form of $k/T$, the $k$ is frequency index and the $T$ is duration of one trajectory that equals to $\Delta t \cdot N_v$.

In a similar manner, the Fourier transform of VAC is written as:

$$VAC(f_k)^{\mathcal{F}} = \Delta t \sum_{j=1}^{N_v} VAC(t_j) \cdot e^{i2\pi f_k t_j} = \Delta t \sum_{j=1}^{N_v} VAC(t_j) \cdot e^{i2\pi k j/N_v}. \tag{6}$$

In addition, according to the definition of the power spectrum as follows:

$$FPS(f_k) = \langle |\boldsymbol{v}_k|^2 \rangle / T, \tag{7}$$

it's easy to derive a concrete expression of $FPS(f_k)$ by plugging the equation (5) into the formula (7), and the final result is given as:

$$FPS(f_k) = \frac{(\Delta t)^2}{T} \sum_{j_1=1}^{N_v} \sum_{j_2=1}^{N_v} \langle \boldsymbol{v}_{j_1} \cdot \boldsymbol{v}_{j_2} \rangle \cdot e^{i2\pi f_k(t_{j_1}-t_{j_2})} = \Delta t \sum_{j=1}^{N_v} VAC(t_j) \cdot e^{i2\pi f_k t_j} = VAC(f_k)^{\mathcal{F}}, \tag{8}$$

where the $j_1$ and $j_2$ are time indexes and the symbol $\mathcal{F}$ represents Fourier transform. So far, one can follow the procedures above to obtain the FPS of migration velocities, and further analyze how the FPS profile changes with frequency on the *log-log* axis.

**Power spectrum of OU-process**

In previously published article [35], Flyvbjerg *et al.* deduced the power spectrum of migration velocities in an OU-process (abbreviated as OUPS), which is given by:

$$OUPS_v(f_k) = OUPS_v^{(true)}(f_k) + \frac{4\sigma_p^2}{\Delta t}[1 - cos(\pi f_k/f_{Nyq})], \tag{9}$$

where the first term on the right-hand side of the equation is the true power spectrum that is not affected by positional errors, and it is written as:

$$OUPS_v^{(true)}(f_k) = \frac{1-c^2}{c}\left(\frac{P}{\Delta t}\right)^2 OUPS_v^{(aliased)}(f_k) + 4D\left(1 - \frac{1-c^2}{c}\frac{P}{\Delta t}\right), \tag{10}$$

in which the aliased term is given as:

$$OUPS_v^{(aliased)}(f_k) = \frac{\langle|\mathbf{v}_k|\rangle^2}{t_{msr}} = \frac{2D(1-c^2)\Delta t/P}{1+c^2-2cos(\pi f_k/f_{Nyq})}. \tag{11}$$

Moreover, the second term denotes the contribution of positional errors. For the equations above, some arguments can be further computed using definite formulas, including Nyquist frequency $f_{Nyq} = 1/(2\Delta t)$ that is a half of the sampling frequency, $c = exp(-\Delta t/P)$, time span $t_{msr} = \Delta t \cdot N_v$ that is identical to $T$, and diffusion coefficient $D = S^2P/2$. It should be noted that the Greek letter $\mathbf{v} = d\mathbf{r}/dt$ represents instantaneous velocity, and it is correlated with secant-approximated velocity $\mathbf{v} = \Delta \mathbf{r}/\Delta t$ by the following integral:

$$\mathbf{v}_i = \int_{t_{i-1}}^{t_i} \mathbf{v}(t')dt'. \tag{12}$$

**Maximum likelihood estimation of motility parameters**

On the basis of the OUPS and FPS above, we can further determine the unknown parameters $\theta = \{D, P, \sigma_p\}$ by using the theoretical OUPS to fit the simulated or experimental FPS. Generally, a routine approach is to first bin-average the FPS values along the frequency axis and then fit the OUPS to these averaged values using a least-squared fitting. However, the least-squared fitting is not optimal because the distribution of these averaged values is not Gaussian [19,35].

In this paper, the solution is to apply maximum likelihood estimation (MLE) to derive the motility parameters $\theta$. Specifically, for a given power spectrum $\{|\mathbf{v}_k|^2/t_{msr}\}_{k=0,\cdots,N_v-1}$, the corresponding log-likelihood function $\ell$ is constructed with the following form:

$$\ell\left(\theta\left|\frac{|\mathbf{v}_k|^2}{t_{msr}}\right.\right) = 2\sum_{k=1}^{N_v} log\left(\frac{2}{OUPS_v(f_k)}\right) + \sum_{k=1}^{N_v} log\left(\frac{|\mathbf{v}_k|^2}{t_{msr}}\right) - \sum_{k=1}^{N_v} log\left(\frac{2}{OUPS_v(f_k)} \cdot \frac{|\mathbf{v}_k|^2}{t_{msr}}\right). \tag{13}$$

As illustrated in equation (9), the theoretical $OUPS_v(f_k)$ is strictly dominated by motility parameters $\theta$, thus one can tune these parameters to maximize the function $\ell_{max}$ taking the $|\mathbf{v}_k|^2/t_{msr}$ as input, and the parameters $\theta_{max}$ is what we want. Finally, derive migration speed $S$ based on the formula of $D = S^2P/2$, and obtain a set of parameters of $\theta'_{max} = \{S, P, \sigma_p\}$.

**WPS**

To investigate the time-varying characteristics of migration velocities in this work, wavelet transform is introduced to compute wavelet power spectrum (WPS), and further derive motility parameters $\theta'_t$ that change over time. Unlike the ensemble-temporal-averaged properties measured by the MSD, VAC, and FPS evaluations, the wavelet transform has demonstrated superior performance in the analysis of the local properties of a non-stationary and infinitely correlated process. Moreover, the wavelet transform also has significant advantages over the windowed Fourier transform, which performs a sliding window of a constant time interval on a time series, *i.e.*, the window size of the former can vary over the frequency. For a given time series of $\mathbf{v}_n$, the wavelet transform is performed by calculating the convolution of $\mathbf{v}_{n'}$ with a wavelet function $\psi_0^*(\eta)$, and this process can be described by:

$$W_n(s) = \sum_{n'=0}^{N_v-1} \mathbf{v}_{n'} \cdot \psi_0^*\left[\frac{(n'-n)\cdot\Delta t}{s}\right], \tag{14}$$

where the $s$ is wavelet scale that correlates with wavelet frequency, and the $\psi_0^*(\eta)$ is a normalized function in which the $\eta$ is a nondimensional "time" parameter, the subscript "0" indicates that the $\psi$ has been normalized, and the asterisk "*" denotes complex conjugate.

Here, *Morlet* is utilized as a wavelet function and it is defined as:
$$\psi_0(\eta) = \pi^{-1/4} \cdot e^{i\omega_0\eta} \cdot e^{-\eta^2/2}, \tag{15}$$
where the $\omega_0$ is a nondimensional frequency and satisfies the admissibility condition when $\omega_0$=6. It is worth noting that the *Morlet* is a complex function. Thus, the convolution result $W_n(s)$ is also complex, consisting of real and imaginary parts. Further, one can easily compute the square of the module, *i.e.*, $|W_n(s)|^2$, and obtain the wavelet power spectrum.

Based on the WPS, we not only understand how the WPS values vary over frequency but also grasp how the characteristics (or modes) encoded in these values evolve with time. Here, the spectrum profile versus frequency at each time point is referred to as the "local" WPS, which vividly shows how the spectral values are distributed, especially the difference between high-frequency and low-frequency domains. It has also been reported that the local WPS is identical to the FPS of the univariate lag-1 autoregressive AR(1) process, on average [33].

In addition, we can analyze in depth the WPS by two averages, one is to average all spectral values along the frequency axis and obtain frequency-averaged results that are usually used to evaluate the migration activity (or energy) at each time point; another is to average all values along the time-axis and obtain time-averaged results that are also called "global" WPS. The global WPS is an unbiased and reliable estimate of the true power spectrum for any time series, and Torrence *et al*. also validated that the global WPS is an approximation to the FPS of the OU process. Hereto, we can continue to fit the theoretical OUPS to the local WPS using the MLE approach and further derive time-varying motility parameters $\theta'_t$, which can be inputted into the PRW model to simulate migration trajectory at each time point.

**Phase diagram**

In this paper, to systematically explore how the motility parameters affect the dynamic metrics, we first simulate a larger number of migration trajectories using the PRW model with a given set of arguments as input, including i) persistence time $P$ increasing from 0.2 to 20 with an increment of 0.2 min, ii) migration speed $S$ increasing from 0.1 to 10 with an increment of 0.1 $\mu$m/min, iii) positional error $\sigma_p$=0.005 $\mu$m, iv) the number $N_p$=5000 of frames (positions) contained in a single trajectory, v) the number $C_n$ of simulated trajectories for any set of arguments, and vi) time step $\Delta t$=0.2 min. Secondly, plot every trajectory in black on a white figure with a size of 1000*1000 pixel$^2$, and mesh the figure uniformly into $N_g$=50*50 grids. Then, count the number $n_c$ of grids covered by a single trajectory and compute the target-finding metric $\alpha$. Meanwhile, compute the distance $r$ from the start to the end of the trajectory and derive migration metric $\beta$ by $r/R$ ($R$=500 pixel). Finally, compute the respective averages of the two metrics based on the $C_n$ trajectories, and draw two contour maps of $\{S, P, \alpha\}$ and $\{S, P, \beta\}$ in OriginLab (2024). It's noted that the motility parameters are obtained for experimental trajectories by fitting the corresponding WPS using the MLE approach rather than prescribing it in advance.

**Heatmap**

Based on the time-varying dynamic metrics for individual trajectories, we continue to explore insights into population dynamics by computing the mean and SD of the two metric series, respectively. The combination of mean and SD is a widely used measure and mainly shows how the values are distributed. In this paper, the two measures are introduced to investigate two aspects, *i.e.*, the stability of the two metrics over time and the difference between the three groups of *D. discoideum*. For any metric series ($\alpha$ or $\beta$) of single-cell trajectory, we can obtain a set of measures $\{mean, SD\}$. Thus, after computing the measures for all migration trajectories, heatmaps can be directly plotted based on kernel density estimation in OriginLab (2024).

### Correlation analysis

To reveal the relationship between motility parameters and dynamic metrics and evaluate the contributions of these parameters to the emergent modes in dynamic metrics, we introduce the correlation coefficient (Pearson, Spearman, Kendall) to quantify the degree of correlation between these parameters and metrics. If the data are continuous numerical variables and satisfy normality, the Pearson correlation coefficient is superior. Otherwise, Spearman or Kendall correlation coefficient is better.

### Author contributions

Y.L. conceived of the presented idea and designed the computational framework; Y.L., D.Q., G.L., and X.L. performed the analytic calculations and the numerical simulations; Y.L. wrote the manuscript with support from Z.L., D.Q., and K.S.; Y.L., Z.L., and W.W. supervised and revised the manuscript. All authors discussed the results and contributed to the final manuscript.

### Competing interests

The authors declare no competing interests.

### Acknowledgements

This work is supported by the National Natural Science Foundation of China (Grant No. 12347178), the Natural Science Foundation of Chongqing, China (Grant Nos. CSTB2022NSCQ-MSX1260), and the Science and Technology Research Program of Chongqing Municipal Education Commission (Grant No. KJQN202300623).

### Data availability

The data that support the findings of this article are available from the previous publication, as described in the Methods section.

### Code availability

The source code in this article is available from the corresponding authors upon request.